\documentclass[9pt,reprint,amsmath,amssymb,pra]{revtex4-1}

\usepackage{graphicx}
\usepackage{bm}% bold math
\begin{document}

\title[Large field enhancement]{Large field enhancement obtained by combining Fabry-Perot resonance and Rayleigh anomaly in photonic crystal slabs}

\author{Kokou B. Dossou}

\altaffiliation{Centre for Ultrahigh-Bandwidth Devices for Optical Systems (CUDOS),
and School of Mathematical and Physical Sciences,
University of Technology Sydney,
PO Box 123,  Broadway,
New South Wales 2007, Australia}

\email{Kokou.Dossou@uts.edu.au}

% \vspace{10pt}
% \begin{indented}
% \item[]\today
% \end{indented}
%

\begin{abstract}
By applying the properties of Fabry-Perot resonance and Rayleigh anomaly, we have shown that
 a photonic crystal slab can scatter the light from an incident plane wave into a diffracted light
with a very large reflection or transmission coefficient.
The enhanced field is either a propagating diffracted wave (with a grazing angle of diffraction)
or a weakly evanescent diffracted wave, so it can be particularly useful for applications requiring an enhanced propagating field
(or an enhanced field with a low attenuation).
An efficient effective medium technique is developed for the design of the resonant photonic crystal slabs.
Numerical simulations have shown that photonic crystal slabs with low index contrast,
such as the ones found in the cell wall of diatoms,
can enhance the intensity of the incident light by four orders of magnitude.
\end{abstract}

% \noindent{\it Keywords}: Photonic crystals, Fabry-Perot resonance, field enhancement, Rayleigh anomaly.

% \submitto{\JOPT}

% \ioptwocol

\maketitle

%
%%%%%%%%%%%%%%%%%%%%%%%
%

%
%%%%%%%%%%%%%%%%%%%%%%%%%%%%%%%%%%%%%%%%
%

\section{Introduction}

The use of diffraction gratings to enhance electromagnetic fields has attracted an extensive research interest for many decades 
because of its rich physics and its potential applications.
The field enhancement is commonly due to a resonant coupling between diffracted plane waves and grating modes~\cite{Hessel:AO:1965}. 
Metallic gratings can support surface modes (surface plasmon polaritons) at metal-dielectric interfaces and the electric field enhancement  near these gratings 
can lead to surface-enhanced Raman scattering~\cite{Mills:PRB:1982,Weber:PRB:1983,Garcia:OC:1983,Garcia:SS:1984,Gao:OE:2009}.
The guided-mode resonance~\cite{Wang:AO:1993,Rosenblatt:IEEE:1997} in dielectric grating waveguide structures
can lead to 
a local enhanced electromagnetic field (which can be useful for enhancing light-matter interactions~\cite{Siltanen:APL:2007}
and fluorescence emission~\cite{Zhang:APL:2008,Laroche:OL:2007}),
a rapid spectral variation
(which has found application in devices such as  optical filters~\cite{Wang:AO:1993})
or an efficient optical reflection by an array of sub-wavelength objects~\cite{Gomez:Medina:OE:2006}.
These resonances typically occur at wavelengths where an evanescent diffraction order is resonantly coupled to a mode 
confined in the diffraction grating,
and so the intensity of the corresponding diffracted wave decreases exponentially as it propagates away from the grating.
The spatial localisation of the evanescent wave means that devices which 
rely on high-field intensity must be operated inside or in the vicinity of the resonant grating.
In some cases (an example will be discussed below), it may be desirable to have a resonant grating where 
the enhanced field remains strong beyond the near-field.
This can be possible if the diffraction order, which is resonantly coupled to a grating mode,
is a propagating order or a weakly evanescent order.

\begin{figure}[htbp]

\centerline{
\includegraphics[width=6cm]{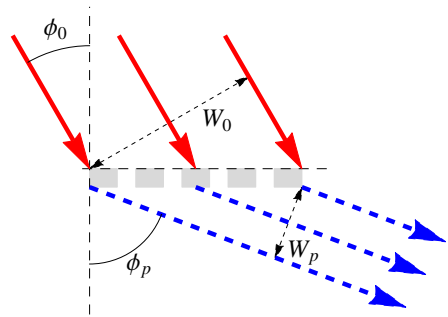}
}

\caption{Spatial compression of an incident plane wave: 
For a grating of length $L$,
the width of the incident plane wave and a diffracted plane wave are respectively
$W_0 = L \cos \phi_0$ and $W_p = L \cos \phi_p$.
There is a spatial compression  of the incident plane wave if $W_p /W_0<1 $, i.e., if $\cos \phi_p / \cos \phi_0 < 1$.}

\label{Fig:spatial:compression}

\end{figure}

\medskip
A weakly evanescent diffraction order can exist near a Rayleigh anomaly (also called Wood anomaly), 
where the transition from a propagating order to an evanescent order takes place.
A propagating diffracted field of order $p \neq 0$ can also have much larger amplitude 
(i.e., a large diffraction coefficient $T_p$)
than the incident wave when it is close to a Rayleigh anomaly.
Indeed, if $\phi_0$ and $\phi_p$ are respectively  the angle of incidence
and the angle of diffraction, the diffraction efficiency $e_p=|T_p|^2 \, \cos \phi_p / \cos \phi_0$
is less than or equal to one, so that we have 
$|T_p|^2 \leq \cos \phi_0 / \cos \phi_p $.
This implies that $|T_p|$ can potentially take large values when $\phi_p$ is near $\pm \pi/2$
(grazing angle of diffraction), which means that the diffraction order is close to a Rayleigh anomaly.
The high-intensity diffracted field, which can occur at a grazing angle of diffraction, can be interpreted as a spatial compression of the incident wave.
This is illustrated in Fig.~\ref{Fig:spatial:compression} where,
for an incidence over a grating of length $L$, the ratio of the width of a diffracted plane wave emerging from the grating 
to the width of the incident plane wave is equal to $\cos \phi_p / \cos \phi_0$.

\medskip
The observations in the previous paragraph suggest that a  propagating diffracted wave (or a weakly evanescent diffracted wave)
with enhanced field can be obtained when a grating waveguide resonance and a Rayleigh anomaly coexist.
In this work, by conceptualising a photonic crystal slab as a perturbation to a uniform dielectric film, 
we show that the properties of the Fabry-Perot resonance can be used to
design photonic crystal slabs that can diffract a substantially enhanced field 
into non-specular orders which are either propagating or weakly evanescent.
The Fabry-Perot model for uniform dielectric films can also be used to accurately predict (especially for photonic crystal slabs with a low index contrast)
the location of the resonance wavelength of a photonic crystal slab.

\medskip
This work is partly motivated by the fact that photonic crystal slabs can be found on the cell wall (frustule)
of some  unicellular photosynthetic organisms known as diatoms~\cite{Fuhrmann:APb:2004}.
The biological function of these periodic photonic structures is not well understood yet~\cite{Fuhrmann:APb:2004,Di:Caprio:JB:2012},
although it has been reported~\cite{Furukawa:Protoplasma:1998} (see also \cite{Fuhrmann:APb:2004,Di:Caprio:JB:2012}) 
that high intensity light can induce 
a movement of the photosynthetic units (chloroplasts) away from the cell wall toward the cell centre
while a weak light can induce a movement toward the cell wall.
This may suggest that the photonic crystal slabs in the diatom frustule can enhance the intensity of an incident light.
In this work we will carry out some numerical simulations of a photonic crystal slab described in~\cite{Fuhrmann:APb:2004},
in order to verify if it can provide an efficient field enhancement.
In such a case, an enhanced field which is propagating or weakly evanescent can be beneficial
since the chloroplasts is not in a direct contact with the frustule, which is located outside the cell membrane.

\medskip
The combination of a Rayleigh anomaly with another optical resonance effect has been previously studied,
especially for the purpose of  improving the transmission efficiency.
The coupling of a Rayleigh anomaly with a surface plasmon polariton~\cite{Gao:OE:2009,McMahon:OE:2007} 
or a Fabry-Perot resonance~\cite{Rahman:OL:2012} has been applied to improve the transmission efficiency through 
metallo-dielectric gratings.
Optical transmission filters with a sharp peak and an improved efficiency have also been demonstrated for dielectric gratings
where a guided-mode resonance and a Rayleigh anomaly coexist~\cite{Amin:APL:2013}.

\medskip
The grating scattering problems in this work have been solved by using a modal method~\cite{Dossou:JOSAA:2012,Sturmberg:CPC:2016},
where the Bloch modes are computed with a finite element method~\cite{Dossou:CMAME:2005}.
We have observed an excellent agreement with the results computed by another finite element-based numerical technique~\cite{Dossou:JCP:2006}.
The modal expansion technique in~\cite{Dossou:JOSAA:2012,Sturmberg:CPC:2016} has some similarity with 
the rigorous coupled wave analysis (RCWA)~\cite{Moharam:JOSAA:1995}
or the Fourier modal method (FMM)~\cite{Li:JOSAA:1997,Popov:PRB:2000}, where the eigenmodes are expressed as Fourier expansions.
We have developed a one-dimensional version of the modal method~\cite{Dossou:JOSAA:2012,Sturmberg:CPC:2016} 
for the numerical simulation of one-dimensional photonic crystal slabs
and an interesting Bloch mode treatment for one-dimensional periodic structures is also presented in~\cite{Xie:OE:2006}.
In many cases, by taking into account the geometry of the problem, an approximate simplified analytical model (which captures the main features of the problem) 
of the modal expansion technique can be developed~\cite{Martin:Moreno:PRL:2001,Garcia:Vidal:PRB:2002,Garcia:Vidal:RMP:2010,Lalanne:JOa:2005}.
These simplified models can provide some useful physical insight and, indeed, we have applied this type of treatment  (effective medium technique) in this work.

\medskip
In what follows, the principles behind the design of resonant photonic crystal slabs will be explained.
In Section~\ref{section:R:T:PCS}, we will give some details on the numerical method used for the calculation of
reflection and transmission through a photonic crystal slab; an effective medium technique will also be presented.
The numerical simulation results of some examples of photonic crystal slabs will be given in the following section.
Finally, in Section~\ref{section:applications}, we will discuss some potential applications of the resonant photonic crystal slabs.
%

%
%%%%%%%%%%%%%%%%%%%%%%%
%

%

%
%%%%%%%%%%%%%%%%%%%%%%%
%

%
%%%%%%%%%%%%%%%%%%%%%%%
%

\begin{figure}[htbp]

\centerline{
\includegraphics[width=5cm]{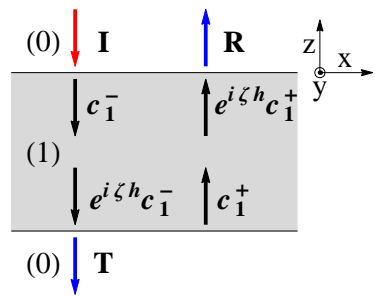}
}

\caption{Illustration of an incidence over a dielectric slab of thickness $h$.
The symbols $I$, $R$ and $T$ represent respectively the incident, reflected and transmitted fields.
The field inside the slab is a superposition of two counter-propagating plane waves 
$c_1^-$ and $c_1^+$, with a propagation constant $\zeta$.}

\label{fig:slab:interfaces}
\end{figure}

%
%%%%%%%%%%%%%%%%%%%%%%%
%

%
%%%%%%%%%%%%%%%%%%%%%%%%%%%%%%%%%%%%%%%%
%

\section{The design principle}

\label{section:design}

The design is based on the following principles. 
First, we consider the problem of plane wave incidence over a uniform dielectric slab of a finite thickness $h$.
Figure~\ref {fig:slab:interfaces} presents an illustration of this problem.
The dielectric slab is denoted (1) and it is surrounded by a medium (0).
The refractive indices of the media (0) and (1) are respectively $n_0$ and $n_1$.
The media (0) and (1) are assumed to be lossless, i.e., $n_0$ and $n_1$ are real numbers; we also suppose that  $n_0 < n_1$.
The wavelength of the incident field is denoted $\lambda$.
The reflection and transmission coefficients $R$ and $T$ through the slab are~\cite{Born:Wolf:1999}:
\begin{eqnarray}
\label{eq1:RT:uniform}
R = \frac{\rho_{01} \left( 1 - e^{2 \, i \, \zeta \, h} \right)}
{ 1 - \rho_{01}^2 \, e^{2 \, i \, \zeta \, h} } 
\quad \mbox{and} \quad
T = \frac{\left( 1 - \rho_{01}^2 \right) e^{ i \, \zeta \, h} } 
{ 1 -  \rho_{01}^2 \, e^{2 \, i \, \zeta \, h} } ,
\end{eqnarray}
where  $\rho_{01}$ is the interface reflection coefficient (or Fresnel reflection coefficient) for incidence from the medium (0)
into the medium (1).
For incidence by a TE-polarised (or $E_y$-polarised) plane wave and a TM-polarised (or $H_y$-polarised) plane wave, 
the mathematical expressions of the reflection coefficient $\rho_{01}$  are respectively~\cite{Born:Wolf:1999}:
\begin{eqnarray}
\label{TE:TM:rho01}
\rho_{01}  = \frac{\gamma - \zeta}{\gamma + \zeta} 
& \mbox{\quad and \quad} &
\rho_{01} = \frac{n_1^2 \, \gamma - n_0^2 \, \zeta}{n_0^2 \, \zeta + n_1^2 \, \gamma} .
\end{eqnarray}
The parameters $\gamma$ and $\zeta$ in Eq.~(\ref{TE:TM:rho01}) are defined as
\begin{eqnarray}
\label{gamma:zeta:Fresnel}
\gamma = \sqrt{n_0^2 \, k_0^2 - \alpha^2} 
& \mbox{\quad and \quad} &
\zeta = \sqrt{n_1^2 \, k_0^2 - \alpha^2} ,
\end{eqnarray}
where $\alpha = n_0 \, k_0 \sin \phi_0$ and $k_0 = 2 \, \pi / \lambda$ is the free space wavenumber.

\medskip
We note that, for both TE and TM polarisations, the interface reflection coefficient $|\rho_{01}|$ 
tends to one as the angle of incidence $\phi_0$ approaches the right angle,
i.e., the interface transmission coefficient $\tau_{01}$ 
tends to zero as the angle of incidence $\phi_0$ approaches the right angle.
This poor interface transmission can be explained by the fact that,  at a grazing angle of incidence, 
the wavevector in the incident medium~(0) is almost tangential to the interface while it can still have a significant normal component
in the transmission medium~(1)
(from Snell's law, the angle of refraction $\phi_1$ converges to $\arcsin (n_0/n_1)$ when the angle of incidence $\phi_0$ approaches the right angle).
By definition, the electric field of a TE-polarised plane wave
and the magnetic field of a TM-polarised plane wave are tangential to the interface.
So when the wavevector in the incident medium is almost tangential to the interface, 
the magnetic field of a TE-polarised plane wave
and the electric field of a TM-polarised plane wave 
are almost perpendicular to the interface, i.e., these fields have a relatively small tangential component.
But both the electric field and the magnetic of field, of the corresponding plane wave in the transmission medium (1),
can still have a significant tangential component.
This field mismatch results in the poor Fresnel transmission at a grazing angle of incidence.

\medskip
However, with the case of plane wave incidence over a dielectric slab of a finite thickness, we can still have a 100\% transmission for a grazing incidence 
when a Fabry-Perot resonance occurs, i.e., when $e^{2 \, i \, \zeta \, h} = 1$ in Eq.~(\ref{eq1:RT:uniform}).
So, at a Fabry-Perot resonance, there is a  high transmission at a grazing incidence,
despite the large field mismatch between the tangential component of a plane wave basis function of the medium~(0)
and the corresponding plane wave basis function in the slab.
As illustrated in Fig.~\ref{fig:slab:interfaces}, the field inside the slab is a superposition of two counter-propagating plane waves.
By using the field continuity equations at the upper and lower slab interfaces,
we can show that the high transmission is due to the fact the pair of counter-propagating plane waves of the homogeneous slab
interact to cancel most of the problematic tangential component 
(i.e., the tangential component of the $\bm{H}$-field and $\bm{E}$-field for respectively the TE and TM polarisations) 
at the two interfaces of the slab, thus allowing a perfect field-matching
with the small tangential components of the incident and transmitted fields.
%

%
%%%%%%%%%%%%%%%%%%%%%%%
%

\begin{figure}[htbp]

\centerline{
\includegraphics[width=7cm]{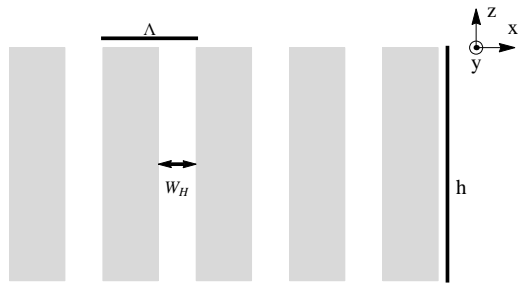}
}

\caption{Illustration of a photonic crystal slab of thickness $h$ and period $\Lambda$.
The photonic crystal consists of one-dimensional array of holes. The width of the holes is denoted $W_H$.}

\label{fig:PC:slab}
\end{figure}

%
%%%%%%%%%%%%%%%%%%%%%%%
%

\medskip
This perfect cancellation effect at a grazing incidence, of two relatively large field components inside the slab,
can make the reflected and transmitted fields very sensitive to a perturbation
of the field inside the slab.
For an example, if the uniform dielectric slab is transformed into a photonic crystal slab (see Fig.~\ref{fig:PC:slab}),
a perfect cancellation (at a Fabry-Perot resonance) cannot be expected 
as the modes in the slab are not plane waves.
As a consequence, when the angle of incidence $\phi_0$ approaches $\pm\pi/2$,
 the perturbation to the slab field  can dominate the small tangential component 
in the incident and transmission media.
In such a situation, a strong contribution from many non-specular diffraction orders can be needed in order 
to match the tangential fields at either side of the slab interfaces,
 and it can be possible that some non-specular propagating orders get exceptionally high diffraction efficiency for a grazing incidence.
We note that the results obtained here can be applied to other forms of perturbation to a uniform dielectric slab.
For an example, a diffraction grating can be placed on top of the uniform dielectric film as in~\cite{Liu:OME:2016}.
The treatment can even be extended to non-uniform layers such as
the photonic crystal gratings in~\cite{Serebryannikov:PRA:2013}, whose top interface is corrugated 
(the photonic crystal in~\cite{Serebryannikov:PRA:2013} is operated in a regime where it behaves approximately as a uniform medium).

\medskip
By reciprocity (see for instance~\cite{Botten:JOSAA:2000,Popov:book}), i.e., reversed incidence, the same photonic crystal slab can also efficiently couple light from an incident plane wave 
into a non-specular plane wave with a grazing angle of diffraction,
and the intensity of the resulting diffracted field can be much higher than the intensity of the incident field
(due to a spatial compression of the incident plane wave).
It follows that the properties, of the Fabry-Perot resonance at a grazing incidence, can be used to design photonic crystal slabs which 
can generate a diffracted wave with a strongly enhanced field.

\medskip
So far we have assumed that the incident field is a propagating incident plane wave, but the results in this section can be extending 
to the case of an incidence by an evanescent plane wave.
The formulas Eqs.~(\ref{eq1:RT:uniform})-(\ref{TE:TM:rho01}) are still valid for an incidence by an evanescent plane wave.
But here the coefficient $\gamma$ in Eq.~(\ref{gamma:zeta:Fresnel}) is a purely imaginary complex number
while $\zeta$ remains a real number (it is assumed that $n_0 \, k_0 < |\alpha| \leq n_1 \, k_0$). 
In particular we have $|\rho_{01}| = 1$, so that the denominator term 
of the reflection and transmission coefficients $R$ and $T$, in Eq.~(\ref{eq1:RT:uniform}),
becomes zero when we have the resonance condition
$2 \, \zeta \, h + 2 \, \arg (\rho_{10})= 2 \, m \pi$, with $m \in \mathbb{N}$.
This resonance condition is often called guided-mode resonance
because the guided modes of a planar waveguide also satisfy the same condition~\cite{Marcuse:book:1974}.
By considering the situation where the purely imaginary complex number $\gamma$ approaches zero,
and by repeating the arguments used for an incidence by a propagating plane wave,
we can design a photonic crystal slab which can enhance the magnitude of a diffracted evanescent field.
%

%
%%%%%%%%%%%%%%%%%%%%%%%%%%%%%%%%%%%%%%%%
%

\section{Reflection and transmission through a photonic crystal slab}

\label{section:R:T:PCS}

We now give some details on the modelling of light transmission through a photonic crystal slab.
We consider the one-dimensional photonic crystal slab shown in Fig.~\ref{fig:PC:slab}.
It has a period $\Lambda$ in the $x$-direction, 
is invariant with respect to $y$ and has a finite thickness $h$ in the $z$-direction.
The photonic crystal slab consists of an array of holes (of width $W_H$ and refractive index $n_0$)
in a medium of refractive index $n_1$ and the refractive index of the upper and lower semi-infinite media is $n_0$.
When a plane wave, with a wavevector $\bm{k}_0 = (\alpha_0, 0, -\gamma_0)$, 
is incident on a photonic crystal slab, the reflected and transmitted fields
can be written as a superposition of plane waves with wavevector
$\bm{k}_p = (\alpha_p, 0, \pm \gamma_p)$,
where $p$ is the diffraction order and the coefficients $\alpha_p$ and $\gamma_p$ are defined as
\begin{eqnarray}
\alpha_p = \alpha_0 + \frac{2 \,  p \, \pi}{\Lambda} 
& \mbox{\quad and \quad} &
\gamma_p = \sqrt{n_0^2 \, k_0^2 - \alpha_p^2} .
\end{eqnarray}
By using the fact that the photonic crystal slab in Fig.~\ref{fig:PC:slab}
can be modelled as a periodic array of $z$-invariant waveguides,
we can write the field inside the slab as a superposition of 
counter-propagating waveguide modes~\cite{Dossou:JOSAA:2012,Sturmberg:CPC:2016}.
This modal expansion is a generalisation of the treatment for a uniform dielectric film,
but here the modal expansion involves a multitude of counter-propagating pairs of modes.
A Bloch mode $\bm{E}_q^{(1)}(x,z)$ of the array of cylinders is quasi-periodic with respect to the $x$ coordinate,
i.e., $\bm{E}_q^{(1)}(x+ \Lambda,z) = e^{i \, \alpha_0 \, \Lambda} \, \bm{E}_q^{(1)}(x,z)$,
and has an exponential dependence $e^{i \, \zeta_q \, z}$ with respect to the $z$ coordinate.
The modes $\bm{E}_q^{(1)}(x,z)$ of the array and their propagation constant $\zeta_q$ can be computed
using a one-dimensional version of the finite element method in~\cite{Dossou:CMAME:2005}.

\medskip
The field continuity conditions at the upper and lower interface of the slab
can be transformed into a system of linear equations~\cite{Dossou:JOSAA:2012} 
by projecting the modal expansion on either side of an interface onto a set of test functions (adjoint modes).
The coefficients of the modal expansions can be obtained by solving this linear system.
The projection involves some overlap integrals between the modes of media (0) and (1).
For an example, let $\Omega$ denotes the upper interface of a unit cell of the array.
If $\Omega$ is at the position $z=0$,
an overlap integral $J_{pq}$ between a plane wave $[ \bm{E}_p^{(0)}(x,z),\bm{H}_p^{(0)}(x,z) ]$ of the medium (0)
and a mode $[ \bm{E}_q^{(1)}(x,z),\bm{H}_q^{(1)}(x,z) ]$ of the medium (1)
can be defined as (see Eq.~(64) of~\cite{Dossou:JOSAA:2012}):
\begin{eqnarray}
\label{overlap:eq1}
J_{pq} & = &
\frac{\displaystyle \int_{\Omega} \left(  \bm{E}_q^{(1)}(-x,0)  \times \bm{H}_p^{(0)} (x,0) \right) \cdot \bm{e}_z \, dx}
{N_p^{(0)} \, N_q^{(1)}} ,
\end{eqnarray}
with
\begin{eqnarray}
N_p^{(0)} & = & \!
\left(\int_{\Omega} \! \left( \! \bm{E}_p^{(0)}(-x,0)  \times \bm{H}_p^{(0)} (x,0) \right) \cdot \bm{e}_z \, dx \! \right)^{1/2} \!\!\! , \\
N_q^{(1)} & = & \!
\left(\int_{\Omega} \! \left( \! \bm{E}_q^{(1)}(-x,0)  \times \bm{H}_q^{(1)} (x,0) \right) \cdot \bm{e}_z \, dx \! \right)^{1/2} \!\!\! .
\end{eqnarray}
For the examples considered in this work, the index contrast of the photonic crystal slab is relatively small and the wave propagation inside the slab 
is typically dominated by a single pair of counter-propagating modes.
If $\zeta_{q'}$  is the propagation constant of the dominant mode, by analogy with a plane wave with a wavevector $\bm{k} = (\alpha_0, 0, \zeta_{q'})$, 
we can define an average refractive index $n_g$ of the photonic crystal slab as
\begin{eqnarray}
\label{average:n}
n_g = \frac{\sqrt{\alpha_0^2 + \zeta_{q'}^2}}{k_0} .
\end{eqnarray}
The value of the average refractive index depends on the wavelength and on the polarisation state.
If, for an incident wavelength $\lambda$, the phase $(\zeta_{q'} \, h)$ is an integer multiple of $\pi$,
a Fabry-Perot resonance will occur for an incidence over a uniform dielectric slab of refractive index $n_g$
and the photonic crystal slab will typically display a resonant behaviour at a wavelength near $\lambda$.
Thus the homogenised slab can be used to predict the approximate location of the resonant wavelengths 
of the photonic crystal slab.

\medskip
For an incidence by a propagating plane wave, by using Eq.~(\ref{average:n}), 
we can express the resonance condition $\zeta_{q'} \, h = m \pi$, with $m \in \mathbb{N}$,  as:
\begin{eqnarray}
\label{resonance:eq1}
\frac{h}{\lambda} \, \sqrt{n_g^2 - n_0^2 \, \sin^2 \phi_0}
& = & \frac{m}{2} .
\end{eqnarray}
Thus for given values of $\lambda$, $\phi_0$, $n_0$ and $n_g$, we can easily find a thickness $h$ corresponding to a resonance.

\medskip
For an incidence by an evanescent plane wave,  the resonance condition
$2 \, \zeta_{q'} \, h + 2 \, \arg (\rho_{10})= 2 \, m \pi$ can be transformed into equations
which are identical to the eigenvalue equation for the guided modes of a planar waveguide~\cite{Wang:AO:1993,Marcuse:book:1974}.
For the TE and TM-polarisations we have respectively
\begin{eqnarray}
\label{disersion:eq1:TE}
\tan (\zeta_{q'} \, h)
& = & \frac{2 \, \hat{\gamma_0} \, \zeta_{q'}}{\zeta_{q'}^2 - \hat{\gamma_0}^2} , \\
\label{disersion:eq1:TM}
\tan (\zeta_{q'} \, h)
& = & \frac{2 \, n_0^2 \, n_g^2 \, \hat{\gamma_0} \, \zeta_{q'}}
{n_0^4 \, \zeta_{q'}^2 - n_g^4  \, \hat{\gamma_0}^2} ,
\end{eqnarray}
with $\hat{\gamma_0} = {\rm Im } \, \gamma_0$.
%

%%%%%%%%%%%%%%%%%%%%%%%%%%%%%%%%%%%%%%%%%%%%%%%

\medskip
For the case of  a grazing angle of incidence, as stated above, other modes, which are different from the ones used in the resonance condition
Eq.~(\ref{resonance:eq1}) or Eqs.~(\ref{disersion:eq1:TE})-(\ref{disersion:eq1:TM}),
can also play a non-negligible role.
As a consequence, the peak reflectance or transmittance wavelength of the photonic crystal slab, 
with a thickness $h$ given by Eq.~(\ref{resonance:eq1})
or Eqs.~(\ref{disersion:eq1:TE})-(\ref{disersion:eq1:TM}),
can be shifted to the left or right side of  the peak reflectance or transmittance wavelength of the uniform dielectric film 
of same thickness $h$ and a refractive index $n_g$
(see Figs.~\ref{Fig:Fabry-Perot:L450:TE} and \ref{Fig:Fabry-Perot:L450:TM} for an example).
As we now explain, the direction of the shift depends essentially on the phase of the photonic crystal modes 
near the resonance wavelength of the uniform slab.
In order to model the plane wave scattering by a photonic crystal slab,
a matrix form of the scalar reflection and transmission coefficients in Eq.~(\ref{eq1:RT:uniform})
must be used and they can be represented as~\cite[see Eqs. (81) and (82)]{Dossou:JOSAA:2012}:
\begin{eqnarray}
\label{eq1:matrix:R}
\bm{\mathcal{R}} & = & \bm{R}_{01} + \bm{T}_{10} \bm{P} (\bm{I}-\bm{R}_{10} \bm{P} \bm{R}_{10}\bm{P})^{-1} \bm{R}_{10} \bm{P} \bm{T}_{01} , \\
\label{eq1:matrix:T}
\bm{\mathcal{T}} & = & \bm{T}_{10} \bm{P} (\bm{I}-\bm{R}_{10} \bm{P} \bm{R}_{10}\bm{P})^{-1}\bm{T}_{01} ,
\end{eqnarray}
where $\bm{P} = {\rm diag \, }[\exp(i \, \zeta_q \, h)]$ is the
diagonal matrix which describes the propagation of the $q^{\rm th}$
Bloch mode inside the photonic crystal slab with a thickness $h$
and $\bm{R}_{01}, \bm{T}_{01}$, $\bm{R}_{10}, \bm{T}_{10}$
are the Fresnel scattering matrices between the medium (0), i.e., the free space
and the medium (1), i.e., the photonic crystal.
In the examples studied in this paper, the wave scattering can be accurately modelled
by  using the plane waves of orders $p=0$ and $q=-1$ (they are the only propagating diffraction orders).
If we truncate the plane wave basis of the medium (0) to the two plane wave functions of diffraction orders 0 and -1,
and select the two Bloch modes of medium (1) which match (have highest overlap) 
these two plane wave functions as basis functions for medium (1), then the matrices in Eqs.~(\ref{eq1:matrix:R}) and (\ref{eq1:matrix:T})
become $2 \times 2$ matrices.
In particular, for incidence from the specular order $p=0$, the reflection coefficient $\mathcal{R}_{0,0}$ 
and the transmission coefficient $\mathcal{T}_{0,0}$ into the order $p=0$ are (here $q$ is set to $q=-1$):
\begin{eqnarray}
\nonumber
\mathcal{R}_{p,p}  = R_{01,p,p} \\
\nonumber
+ \frac{R_{10,p,p} \, T_{01,p,p} \, T_{10,p,p} 
\! \left(1 - R_{10,q,q}^2 \, e^{ 2 \, i \, \zeta_q \, h}  \right) \!  e^{2 \, i \, \zeta_p \, h}   + \delta_N}
{ \left( 1 - R_{10,p,p}^2 \, e^{2 \, i \, \zeta_p \, h} \right)
\left(1 - R_{10,q,q}^2 \, e^{ 2 \, i \, \zeta_q \, h}  \right) + \delta_D}  ,
\\
\label{eq1:scalar:RT:PC:slab}
\\
\nonumber
\mathcal{T}_{p,p} = 
\frac{T_{01,p,p} \, T_{10,p,p} \, 
\left(1 - R_{10,q,q}^2 \, e^{ 2 \, i \, \zeta_q \, h}  \right) 
e^{ i \, \zeta_p \, h} + \delta_N'} 
{ \left( 1 - R_{10,p,p}^2 \, e^{2 \, i \, \zeta_p \, h} \right)
\left(1 - R_{10,q,q}^2 \, e^{ 2 \, i \, \zeta_q \, h}  \right)  + \delta_D} ,
\end{eqnarray}
The terms $\delta_N$, $\delta_N'$ and $\delta_D$ which appear at the numerators or the denominators of Eq.~(\ref{eq1:scalar:RT:PC:slab})
are null when the off-diagonal coefficients of the scattering matrices 
$\bm{R}_{01}, \bm{T}_{01}$,
$\bm{R}_{10}, \bm{T}_{10}$
are zeros, in such a case we can verify that Eq.~(\ref{eq1:scalar:RT:PC:slab}) is equivalent to Eq.~(\ref{eq1:RT:uniform}).
Otherwise, the location of the maximum in reflectance or transmittance is largely determined by the denominator
of $\mathcal{R}_{p,p}$ and $\mathcal{T}_{p,p}$.
Here we have 
\begin{eqnarray}
\delta_D = - R_{10,p,q} \, R_{10,q,p}  \, e^{ i \, (\zeta_p + \zeta_q) \, h} \\
\nonumber
\left(
2 + \left( 2 \,  R_{10,p,p} \, R_{10,q,q} -  R_{10,p,q} \, R_{10,q,p} \rule{0mm}{4mm} \right) 
e^{ i \, (\zeta_p + \zeta_q) \, h}
\right) .
\end{eqnarray}
If the value of $\delta_D$ is a real number near the resonance wavelength of the  uniform dielectric slab,
we have observed that the resonance wavelength of the photonic crystal slab will closely match
that of the uniform slab.
Otherwise, depending on the phase of  $\delta_D$, the resonance wavelength of the photonic crystal slab moves to the left or right side
of the  resonance wavelength of the  uniform dielectric slab.
%

%
%%%%%%%%%%%%%%%%%%%%%%%%%%%%%%%%%%%%%%%%
%

\section{Numerical simulations}

\label{section:simulations}

For the numerical calculations, we first consider the example of a one-dimensional photonic crystal slab which has same lattice constant and filling ratio 
as a two-dimensional square array of holes described in~\cite{Fuhrmann:APb:2004}, 
where the photonic crystal structures found in the cell wall (frustule) of the marine diatom \emph{Coscinodiscus granii}
have been analysed.
We will also analyse a two-dimensional photonic crystal slab later in this section.
The lattice constant of the two-dimensional square array is $\Lambda = 250~{\rm nm}$ and the hole radius is $a = 90~{\rm nm}$.
The slab is made of silica, with a refractive index $n_1=1.43$. 
The slab is surrounded by water and the holes are filled with water (refractive index $n_0 = 1.33$).
The filling ratio of the two-dimensional array of holes with a square lattice is $f_{2D} = \pi \, a^2 /\Lambda^2 = 0.40715 $.
The filling ratio of the corresponding one-dimensional photonic crystal (see Fig.~\ref{fig:PC:slab}) is  $f_{1D} = W_H / \Lambda$,
where $W_H$ is the width of the holes.
We have $f_{1D} = f_{2D}$ if $W_H = 101.788~{\rm nm}$.
For the numerical simulations, we have rounded the value of $W_H$
to $W_H = 101.8~{\rm nm}$.

\medskip
For an illustration of the derivations presented in Section~\ref{section:R:T:PCS}, we want to find a thickness $h$ of the photonic crystal slab
such that a Fabry-Perot resonance occurs at a grazing angle of incidence (e.g., $\phi_0 = 89.999^{\circ}$) 
near the wavelength $\lambda = 450~{\rm nm}$ (blue colour spectrum).
This value of the  wavelength is chosen because diatoms can display
high photosynthetic activity in the blue and red regions of the spectrum~\cite{Kirk:book}.
It is also said in~\cite{Fuhrmann:APb:2004} that the diatom frustule slab, with a square lattice array, has a thickness between 200 and 600~nm,
and, as we shall see, this range includes the slab thickness where a Fabry-Perot resonance occurs at the wavelength $\lambda = 450~{\rm nm}$.
Note however that we have observed the behaviours described below at other wavelengths.
For an incidence by a TE-polarised plane wave,
we have found that the propagation constant $\zeta_{q'}$ of the Bloch mode of the periodic array, which has the highest overlap Eq.~(\ref{overlap:eq1})
with the incident plane wave, is $\zeta_{q'}/k_0 = 0.40036$.
Since $\alpha_0/k_0 = n_0 \, \sin \phi_0 \approx n_0 \, \sin (\pi/2) = n_0 = 1.33$, 
according to Eq.~(\ref{average:n}), the average refractive index (at the wavelength $\lambda = 450~{\rm nm}$) of the photonic crystal is $n_g = 1.3889$.
We can then obtain the resonance thickness from Eq.~(\ref{resonance:eq1}): 
$h \approx 0.5 \, \lambda \, (n_g^2 - n_0^2 )^{-1/2} = 562.2~{\rm nm}$, where the parameter $m$ in Eq.~(\ref{resonance:eq1}) is set to $m=1$.
With the case of an incidence by a TM-polarised plane wave, 
we have $\zeta_{q'}/k_0 = 0.4006$, $n_g = 1.3890$,
so that the computed resonance thickness is $h = 561.7~{\rm nm}$.
We have used the thickness value $h = 562.0~{\rm nm}$ for the calculations.

\begin{figure}[htbp]

\centerline{
\includegraphics[width=7cm]{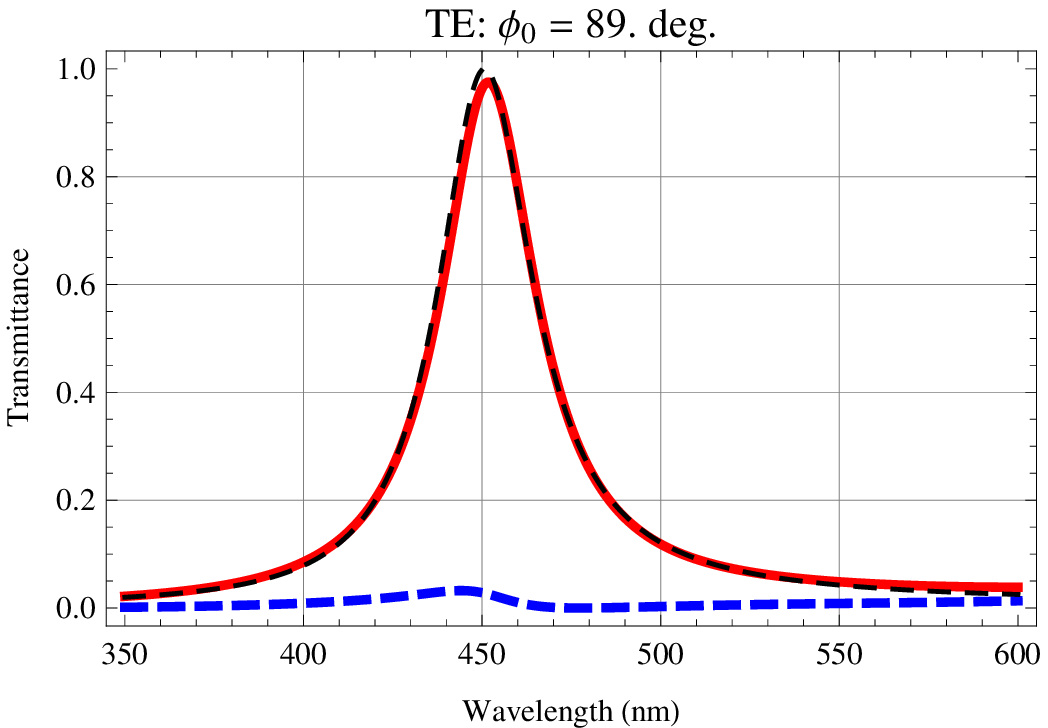}
}
\centerline{
\includegraphics[width=7cm]{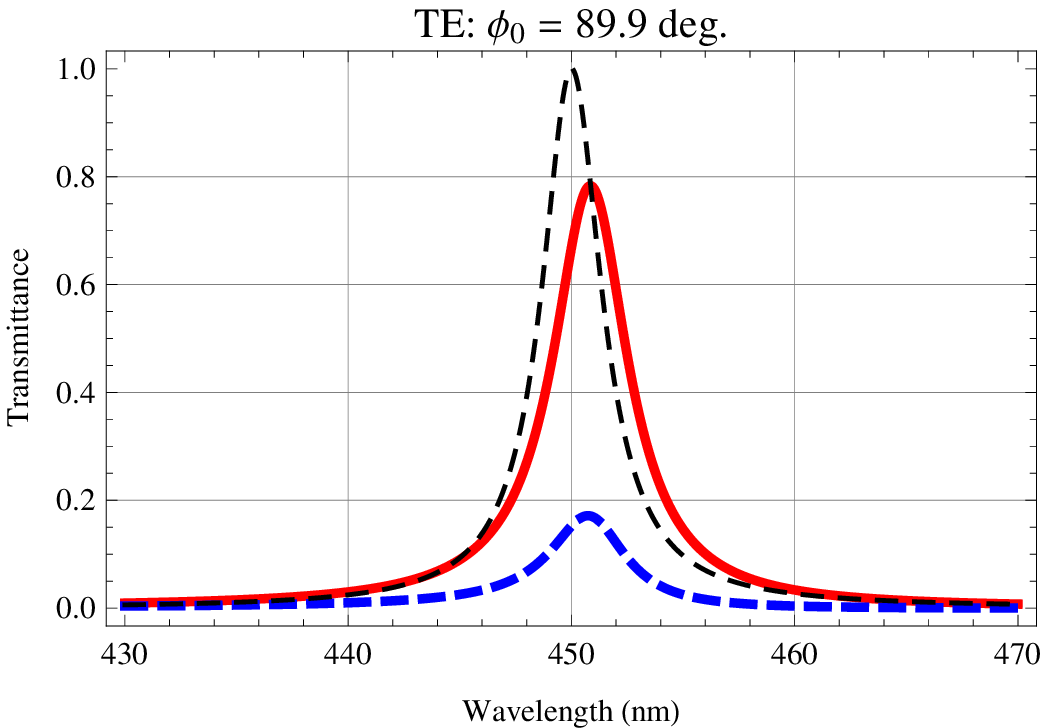}
}
\centerline{
\includegraphics[width=7cm]{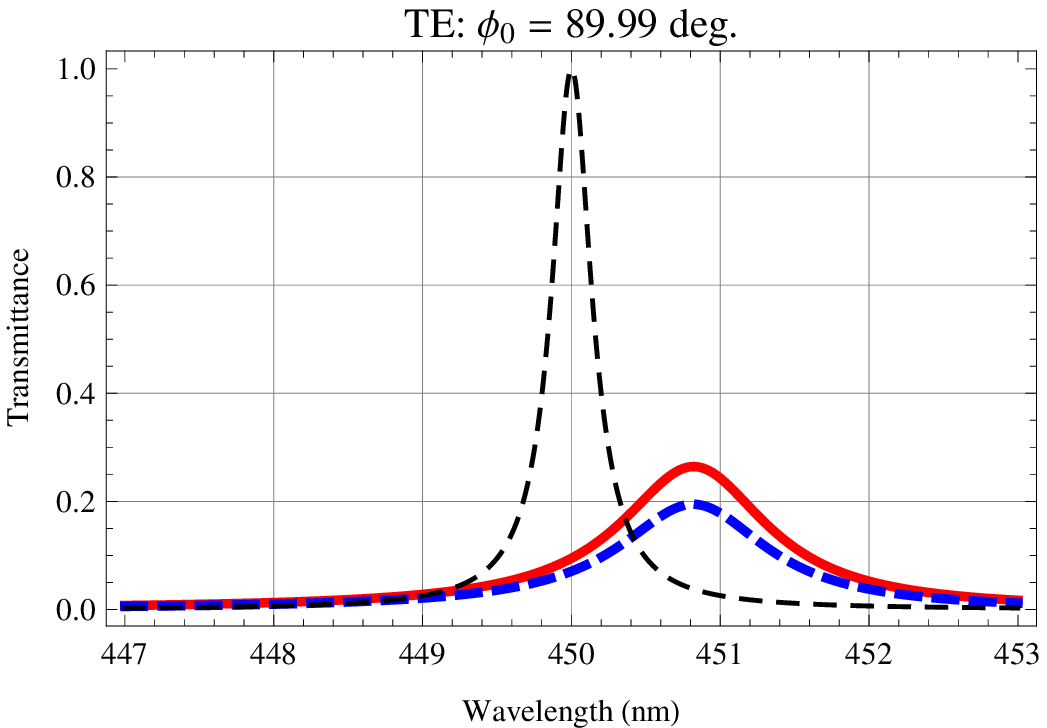}
}
\centerline{
\includegraphics[width=7cm]{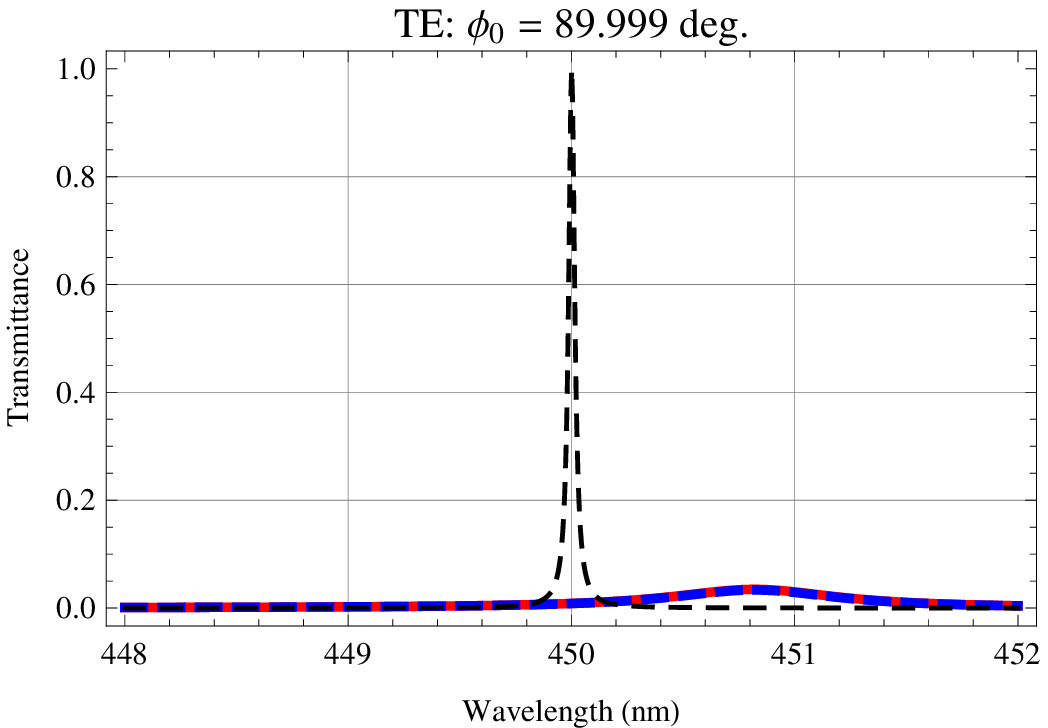}
}

\caption{Fabry-Perot resonance at the grazing incidence angles
$\phi_0=89^{\circ}$, $89.9^{\circ}$, $89.99^{\circ}$ and $89.999^{\circ}$,
near the wavelength $\lambda = 450~{\rm nm}$  (TE-polarisation).
The thick continuous red curves and the thick dashed blue curves show respectively
the total transmittance  
and the diffraction efficiency into the order $p = -1$ of a photonic crystal slab
with a thickness $h = 562.0~{\rm nm}$.
The thin dashed black curves show the transmittance through
a homogeneous slab of  thickness $h$ and refractive index  $n_g(\lambda)$.}

\label{Fig:Fabry-Perot:L450:TE}

\end{figure}

\begin{figure}[htbp]

\centerline{
\includegraphics[width=7cm]{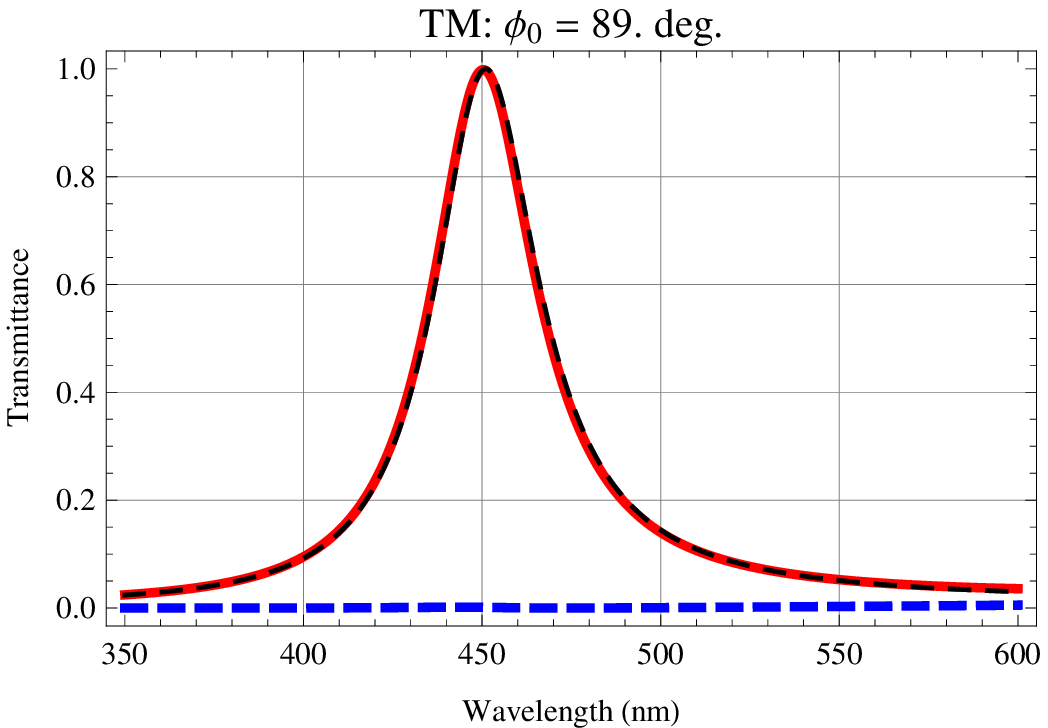}
}
\centerline{
\includegraphics[width=7cm]{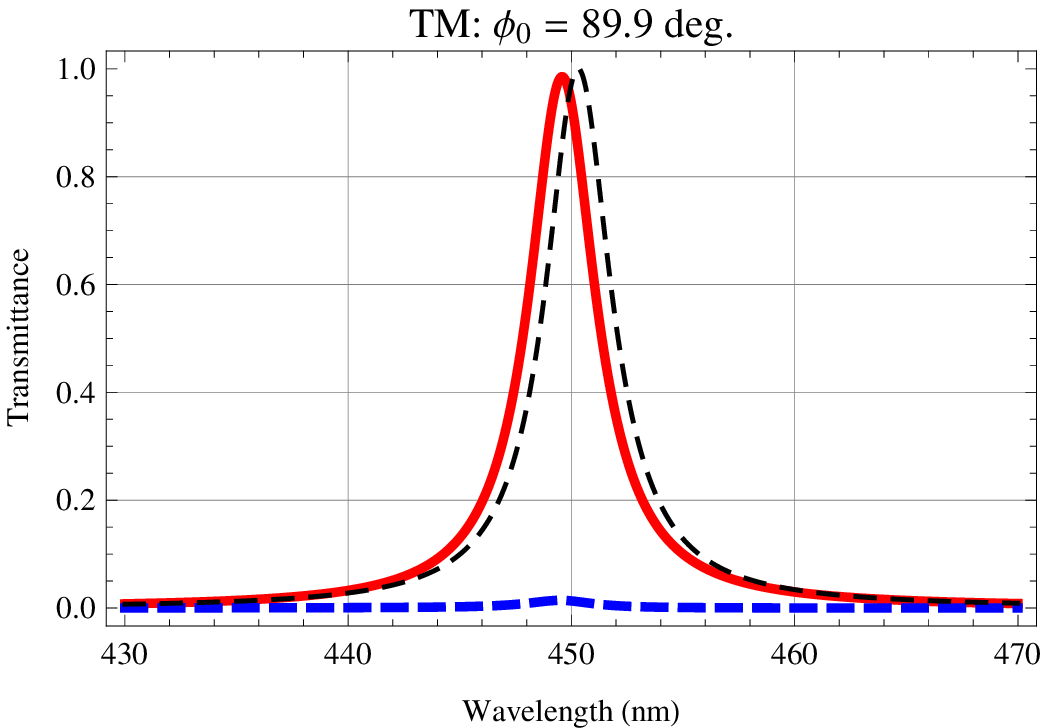}
}
\centerline{
\includegraphics[width=7cm]{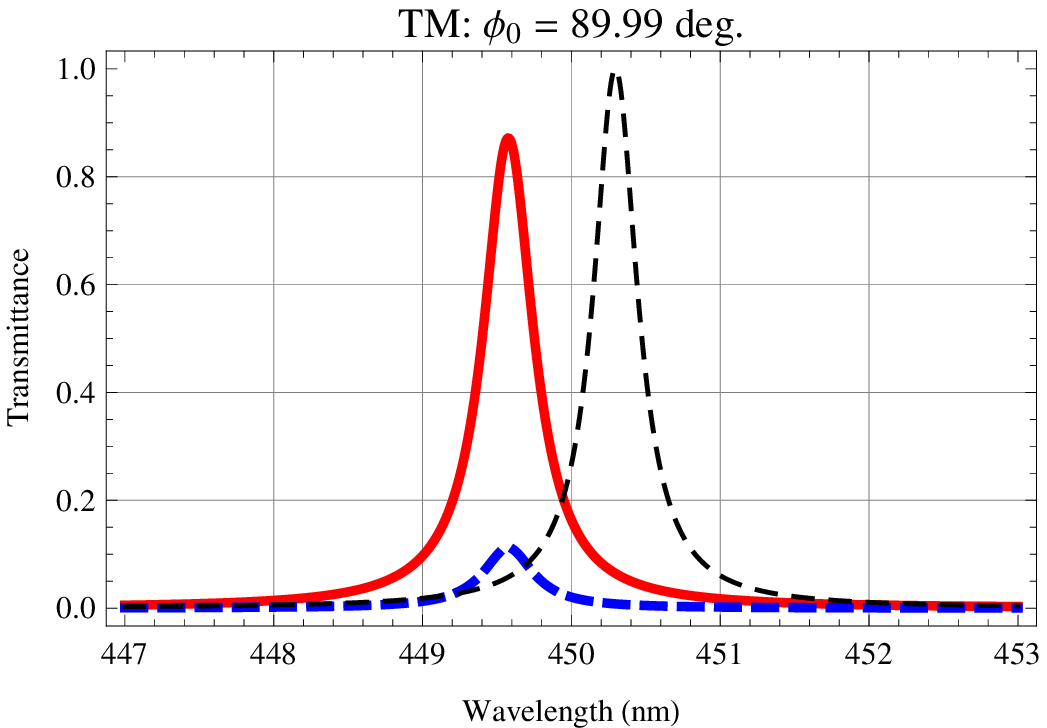}
}
\centerline{
\includegraphics[width=7cm]{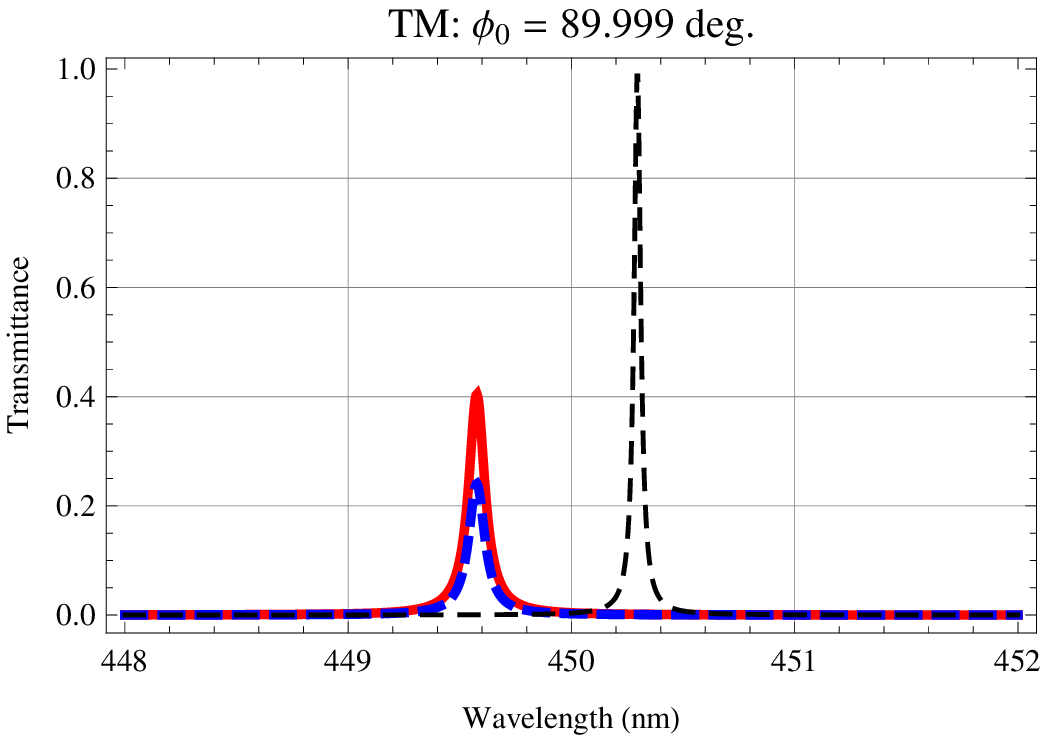}
}

\caption{Fabry-Perot resonance at grazing incidence angles
near the wavelength $\lambda = 450~{\rm nm}$ (TM-polarisation).
Same notation convention as in Fig.~\ref{Fig:Fabry-Perot:L450:TE}.}

\label{Fig:Fabry-Perot:L450:TM}

\end{figure}

\medskip
Figure~\ref{Fig:Fabry-Perot:L450:TE} shows the transmittance through the photonic crystal slab of thickness $h =  562.0~{\rm nm}$ for the angles of incidence
$\phi_0=89^{\circ}$, $89.9^{\circ}$, $89.99^{\circ}$ and $89.999^{\circ}$, for TE-polarisation.
The corresponding results for the TM-polarisation are plotted in Fig.~\ref{Fig:Fabry-Perot:L450:TM}.
In Figs.~\ref{Fig:Fabry-Perot:L450:TE} and \ref{Fig:Fabry-Perot:L450:TM},
the thick continuous red curves and the thick dashed blue curves represent respectively the total transmittance and the diffraction efficiency (in transmission) into the order $p=-1$.
The thin dashed black curves are the transmittance $|T|^2$ (the transmission coefficient $T$ is given by Eq.~(\ref{eq1:RT:uniform})) 
through a uniform slab of refractive index $n_g(\lambda)$ and same thickness as the photonic crystal slab.
In all cases, the transmission peak occurs at a wavelength which is close to the design wavelength $\lambda = 450~{\rm nm}$.
The transmittance of the photonic crystal slab is very close to that of the uniform slab at the  angle of incidence $\phi_0=89^{\circ}$.
But when $\phi_0$ approaches the right angle, as predicted in Section~\ref{section:design}, 
the response of the photonic crystal slab deviates from that of the uniform slab. In particular,
the proportion of power carried by the first diffraction order $p=-1$ starts to increase substantially.
As explained in Section~\ref{section:design}, by reversing the direction of the diffracted plane wave, 
we can also expect an efficient transmission into a grazing angle of diffraction.
%

%
%%%%%%%%%%%%%%%%%%%%%%%
%

%

%
%%%%%%%%%%%%%%%%%%%%%%%
%

%

%
%%%%%%%%%%%%%%%%%%%%%%%
%

\begin{figure}

\centerline{
\includegraphics[width=3.5cm]{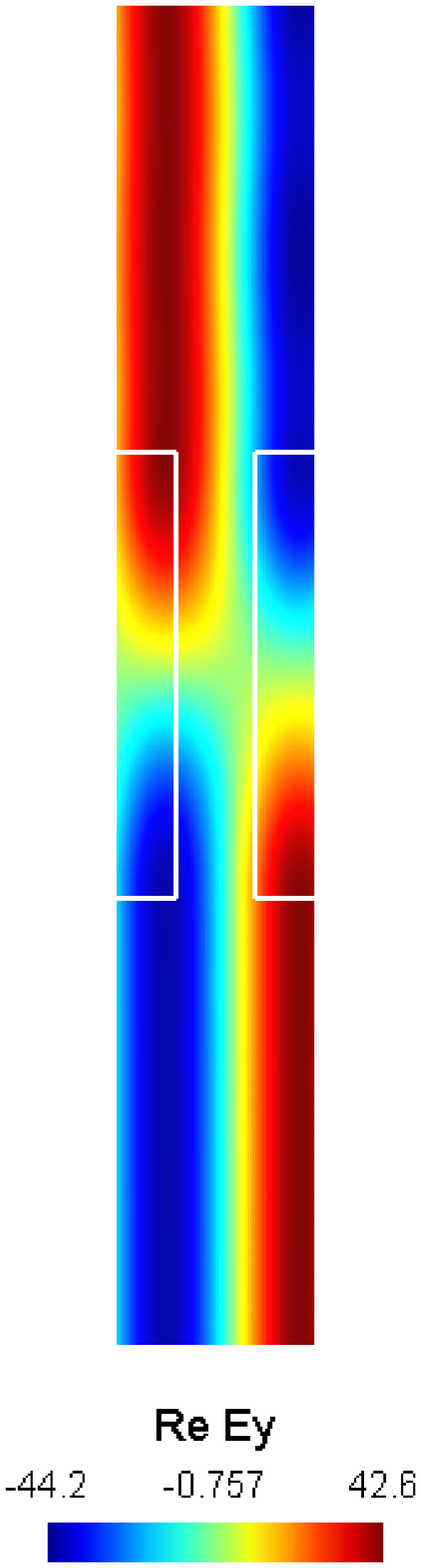}
\includegraphics[width=3.5cm]{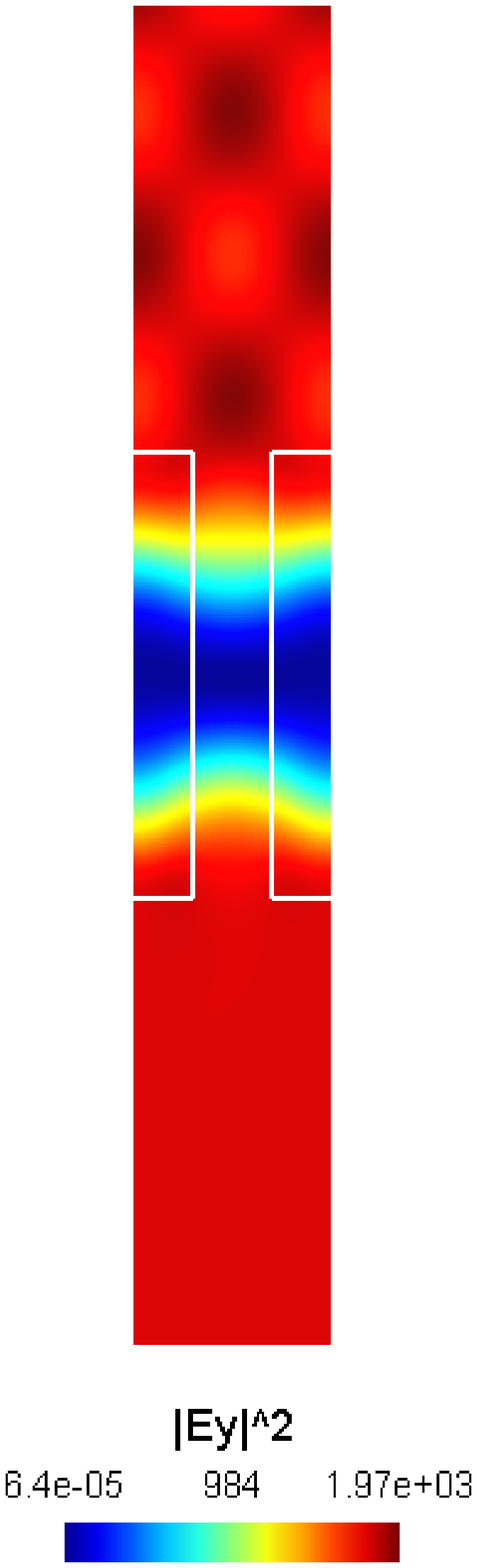}
}

\caption{Plot of the fields ${\rm Re} \, E_y$ and $|E_y|^2$ at the resonance wavelength 450.8~nm for an incidence 
by a TE-polarised plane wave with a unit amplitude
(the angle of incidence is $\phi_0  = 20.845^{\circ}$).}

\label{Field:plot:TE}
\end{figure}

\begin{figure}

\centerline{
\includegraphics[width=3.5cm]{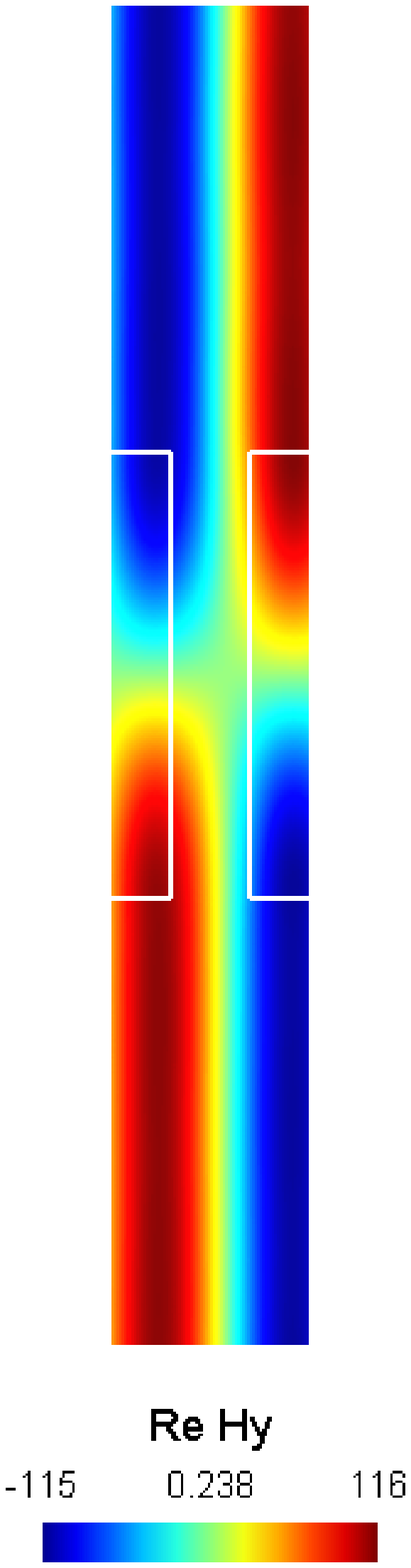}
\includegraphics[width=3.5cm]{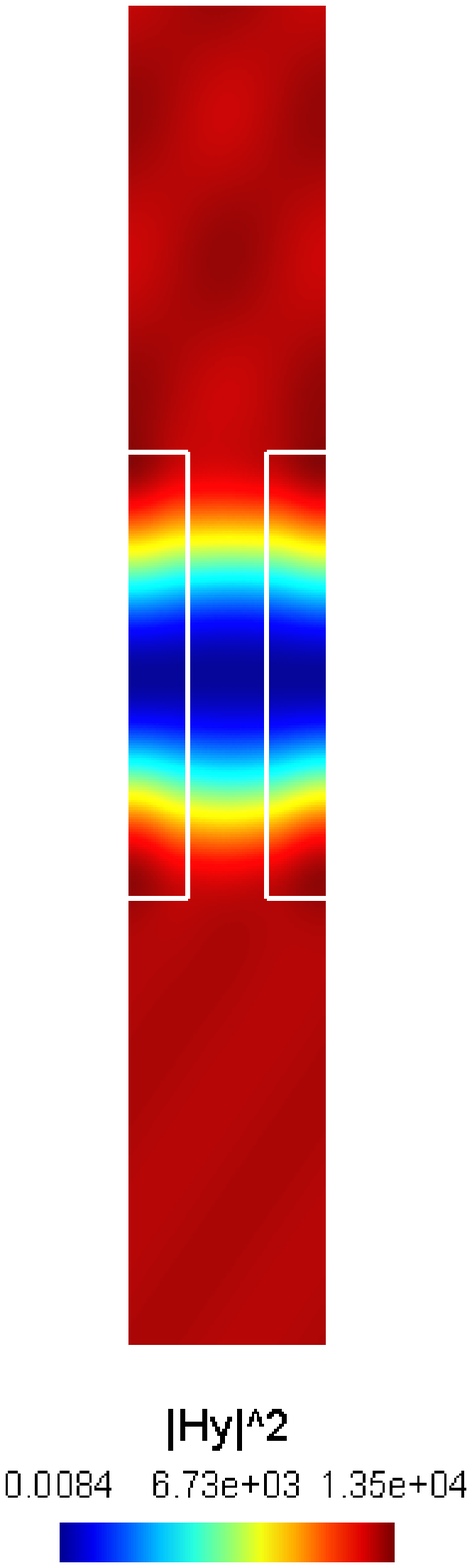}
}

\caption{Plot of the fields ${\rm Re} \, H_y$ and $|H_y|^2$  at the resonance wavelength 449.6~nm for an incidence 
by a TM-polarised plane wave with a unit amplitude
(the angle of incidence is $\phi_0  = 20.617^{\circ}$).}

\label{Field:plot:TM}
\end{figure}

\medskip
Let us show an example of field enhancement.
In Figs.~\ref{Fig:Fabry-Perot:L450:TE} and \ref{Fig:Fabry-Perot:L450:TM}, the peak diffraction efficiency $e_{-1}$ at the incidence angle
$\phi_0 = 89.999^{\circ}$ is $e_{-1} = 0.0335$ (at the wavelength 450.8~nm) for the TE-polarisation 
and $e_{-1} = 0.241$ (at the wavelength 449.6~nm) for the TM-polarisation.
The angle of diffraction at the peak diffraction efficiency is $\phi_{-1} = -20.845^{\circ}$ for TE-polarisation and $\phi_{-1} = -20.617^{\circ}$
for the TM-polarisation.
We recall that the angle of diffraction into an order $p$ is $\phi_{p} = \arcsin (\sin \phi_0 + p \, \lambda/(n_0 \, \Lambda))$.
The transmission coefficients $|T_{-1}| = \sqrt{e_{-1} \, \cos \phi_0 / \cos \phi_{-1}}$ 
for the TE and TM-polarisations are, respectively, 
$|T_{-1}| = 0.0008$ and $|T_{-1}| = 0.0021$.
For the reversed incidence, by reciprocity, the diffraction efficiency $e_{-1}$ is unchanged but the transmission coefficient becomes
$|T_{-1}| = \sqrt{e_{-1} \, \cos \phi_{-1} / \cos \phi_0}$ 
and their values are relatively large:
$|T_{-1}| = 42.35$ for the TE-polarisation
(the corresponding field plots of ${\rm Re} \, E_y$ and $|E_y|^2$ are shown in Fig.~\ref{Field:plot:TE})
and $|T_{-1}| = 113.62$ for the TM-polarisation (the corresponding field plots of ${\rm Re} \, H_y$ and $|H_y|^2$ are shown in Fig.~\ref{Field:plot:TM}).
The field plots in Figs.~\ref{Field:plot:TE} and \ref{Field:plot:TM} show respectively the total field components
 $E_y$ (for an incidence by a TE-polarised plane wave with a unit amplitude $\bm{E} = [0,1,0]$) 
and $H_y$ (for an incidence by a TM-polarised plane wave with a unit amplitude $\bm{H} = [0,1,0]$).
In Figs.~\ref{Field:plot:TE} and \ref{Field:plot:TM}, 
the white lines indicate the interfaces (over a unit cell) between the background material (silica) and the surrounding medium (water).
%
% The green arrows in Figs.~\ref{Field:plot:TE} and \ref{Field:plot:TM} represent the propagation direction of the incoming (incident) plane wave 
% and the outgoing (diffracted) plane waves of orders $p=0$ and $p=-1$.
%
Above the slab unit cells, the periodic features, which can be seen in the field plots of $|E_y|^2$ and $|H_y|^2$,
are caused by the interference between the incident plane wave and the diffracted plane waves of orders 0 and -1.
Below the unit cell, the interference between the transmitted plane waves of orders 0 and -1 is weaker 
because the transmission coefficient into the specular order has a relatively small value.
The results presented here are an interesting confirmation that a photonic crystal slab can be designed to increase the intensity of an incident plane wave 
by an extremely large factor (here we have an example where the factor $|T_{-1}|^2$ is above $10^4$).
This is also consistent with the results in \cite{Amin:APL:2013}, in particular the field plot in Fig.~3(c) of \cite{Amin:APL:2013} 
shows that the orders $p=1$ and $p=-1$ are propagating at a high grazing angle
and the amplitude of their field is about 10 times larger than the amplitude of the incident plane wave.
A giant field enhancement factor of about $10^6$ in photonic crystal slab is also reported in \cite{Mocella:PRB:2015}, again, for a wavelength near a Rayleigh anomaly.
%

%
%%%%%%%%%%%%%%%%%%%%%%%
%

% The green arrows indicate the direction of propagation of the incident and diffracted waves

%
%%%%%%%%%%%%%%%%%%%%%%%
%

%
%%%%%%%%%%%%%%%%%%%%%%%
%

\begin{figure}[htbp]

\centerline{
\includegraphics[width=6.5cm]{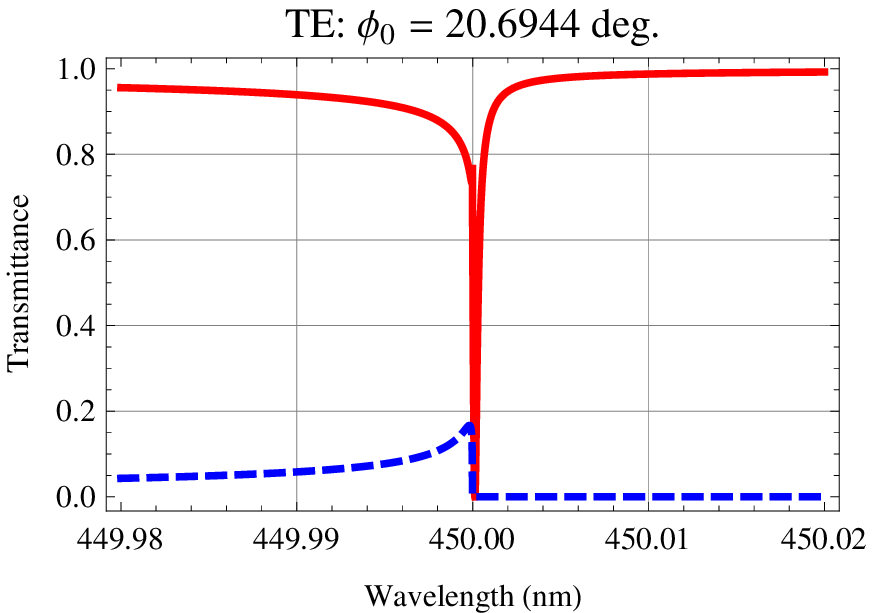}
}

\centerline{
\includegraphics[width=6.5cm]{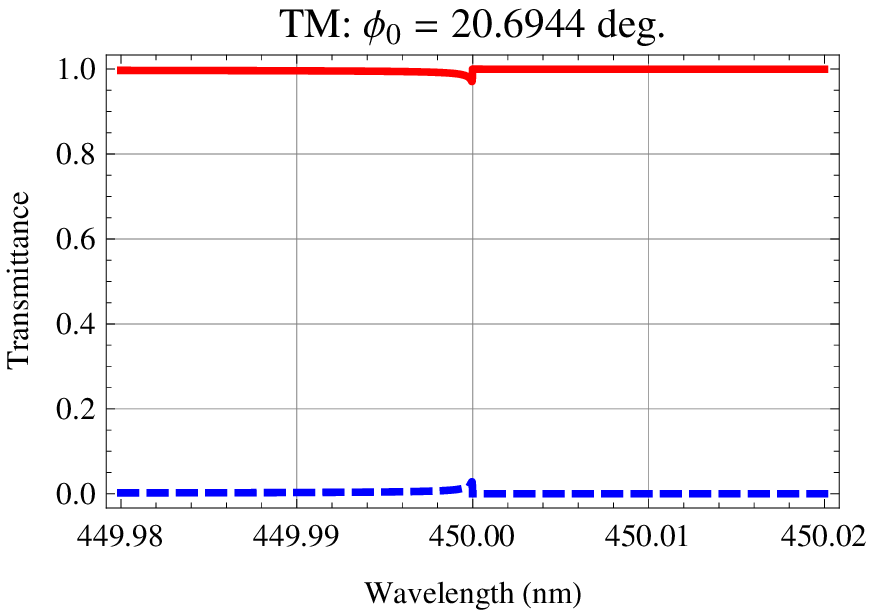}
}

\caption{Fabry-Perot resonance: 
The total transmittance (continuous red curves) and the diffraction efficiency into the order $p=-1$ 
(dashed blue curves)
around the wavelength $\lambda = 450.0~{\rm nm}$, where a Rayleigh anomaly occurs.
The slab thickness is $h = 562.0~{\rm nm}$.}

\label{Fig3:Fabry-Perot:L450}

\end{figure}

\begin{figure}[htbp]

\centerline{
\includegraphics[width=6.0cm]{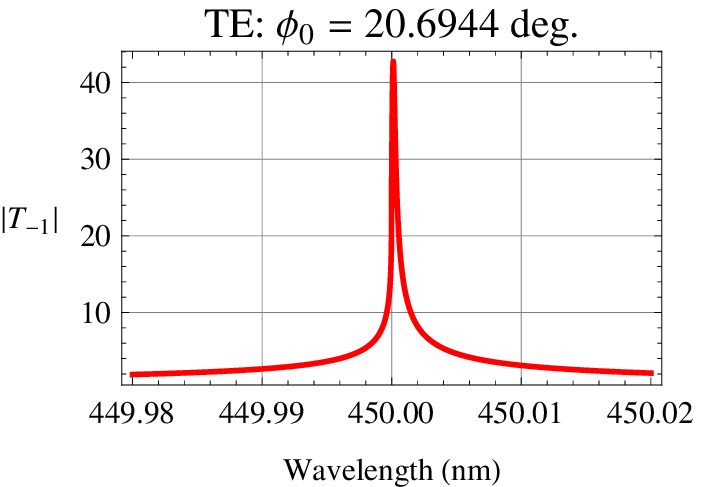}
}

\centerline{
\includegraphics[width=6.0cm]{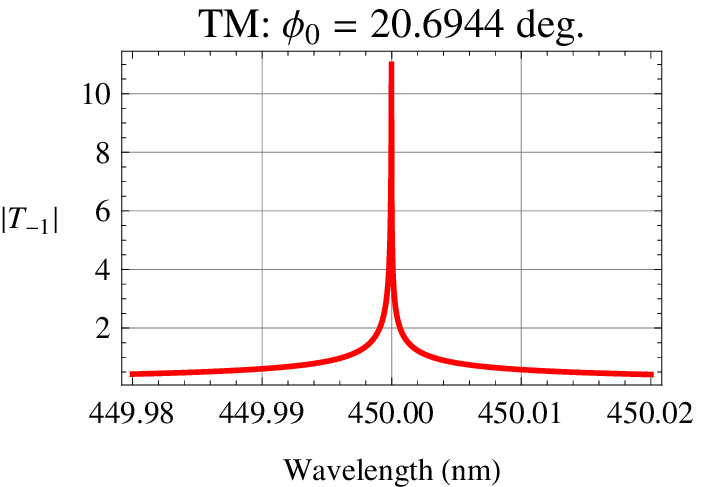}
}

\caption{Field enhancement: Transmission coefficient $|T_{-1}|$  corresponding to the diffraction efficiency curves in Fig.~\ref{Fig3:Fabry-Perot:L450}.}

\label{Fig4:Fabry-Perot:L450}

\end{figure}

\medskip
With the reversed case, the angle of incidence depends on the wavelength.
However, in order to study the wavelength dependence of the quantities $e_{-1}$ and $|T_{-1}|$,
it would be convenient to analyse a problem with a fixed angle of incidence.
Here we will use a reversed incident angle associated with the design wavelength $\lambda = 450~{\rm nm}$:
the angle of diffraction $\phi_{-1}$ at a high grazing angle $\phi_0$ can be approximated as
$\phi_{-1} = \arcsin (\sin (89.999^{\circ}) - \lambda/(n_0 \, \Lambda))  = -20.6944^{\circ}$.
The curves of the total transmittance  (continuous red curves) and the diffraction efficiency $e_{-1}$  (dashed blue curves)
at the angle of incidence $\phi_0 = 20.6944^{\circ}$ are plotted in Fig.~\ref{Fig3:Fabry-Perot:L450}.
The transmission coefficient $|T_{-1}|$ corresponding to the diffraction efficiency $e_{-1}$  is shown in Fig.~\ref{Fig4:Fabry-Perot:L450}.
The angle of incidence $\phi_0$ is chosen such that the Rayleigh anomaly for the order $p = -1$ occurs near the wavelength $\lambda = 450~{\rm nm}$
and the slab thickness $h$ is chosen such that the order $p = -1$ is in a Fabry-Perot resonance with a mode of the photonic crystal slab.
The results in Fig.~\ref{Fig4:Fabry-Perot:L450} indicate that, with a Fabry-Perot resonance near a Rayleigh anomaly, 
a photonic crystal slab can generate a grazing propagating diffracted field
(for $\lambda < 450~{\rm nm}$)
or a weakly evanescent diffracted field (for $\lambda > 450~{\rm nm}$)
with an amplitude that is much higher than the amplitude of the incident light.
%

%
%%%%%%%%%%%%%%%%%%%%%%%
%

%

%
%%%%%%%%%%%%%%%%%%%%%%%
%

%
%%%%%%%%%%%%%%%%%%%%%%%
%

%
%%%%%%%%%%%%%%%%%%%%%%%
%

\begin{figure}[htbp]

\centerline{
\includegraphics[width=6.0cm]{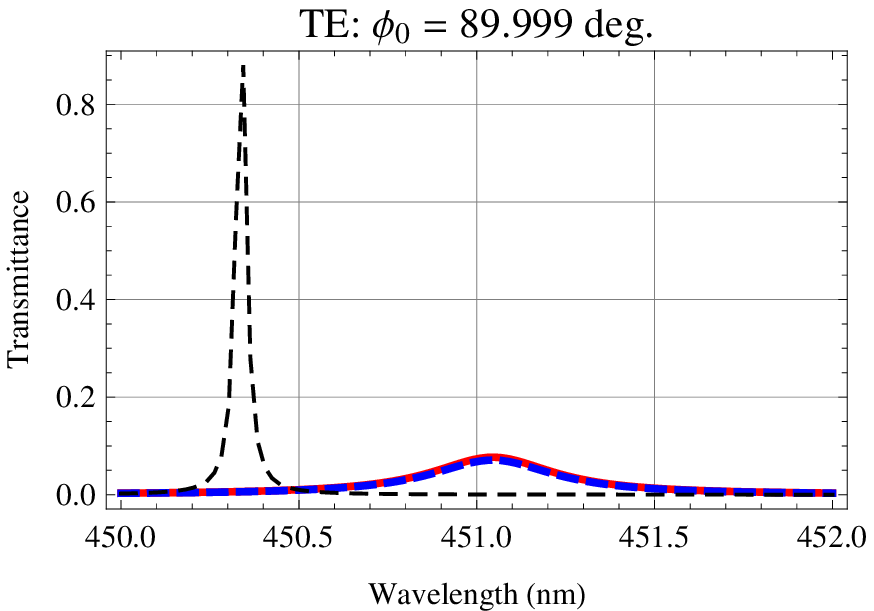}
}

\centerline{
\includegraphics[width=6.0cm]{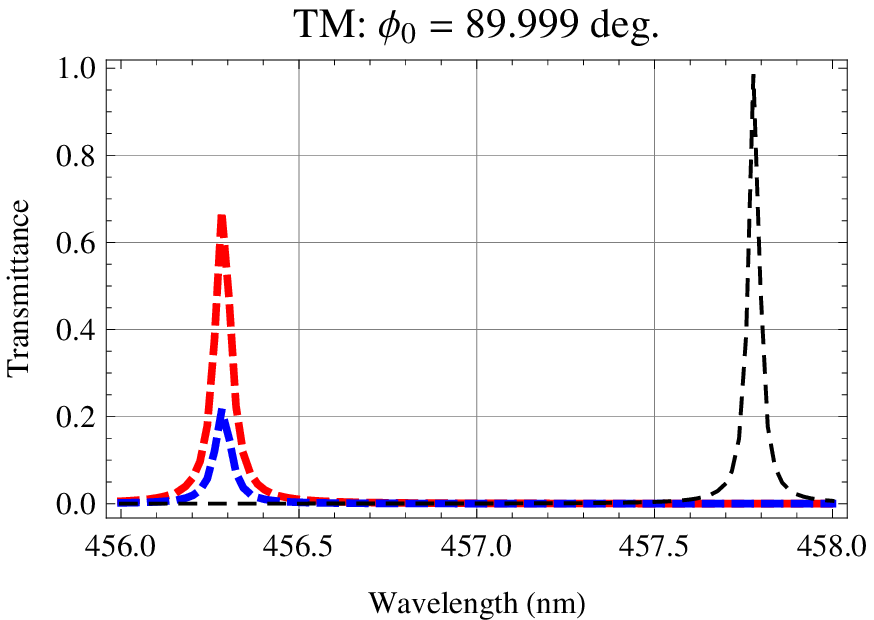}
}

\caption{Two-dimensional photonic crystal slab (Fabry-Perot resonance  at grazing incidence angles):
The thickness of the slab is $h = 563.0~{\rm nm}$.
Same notation convention as in Fig.~\ref{Fig:Fabry-Perot:L450:TE}.}

\label{Fig:Fabry-Perot:L450:2d}

\end{figure}
%

%
%%%%%%%%%%%%%%%%%%%%%%%
%

\medskip
Finally, we also analyse a case of a photonic crystal slab with the two-dimensional square lattice array of holes~\cite{Fuhrmann:APb:2004}
whose parameters were given earlier in this section.
At the wavelength $\lambda = 450~{\rm nm}$, the average refractive index $n_g$ for the TE and TM-polarisations are, respectively, 
$n_g = 1.3888$ and $n_g = 1.3908$.
For the homogenised dielectric slab, the Fabry-Perot resonance at high grazing angle of incidence occurs at the slab thickness
$h = 562.6~{\rm nm}$ (TE-polarisation) and $h = 553.2~{\rm nm}$ (TM-polarisation).
The relative difference between these two thickness values is about 1.6\% (in the case of the one-dimensional photonic crystal slab, the relative difference is 0.065\%) 
and so the two-dimensional photonic crystal slab can be expected to exhibit a higher sensitivity to polarisation.
For the numerical calculation, the slab thickness is set to $h = 563~{\rm nm}$, which is close to the resonance thickness for the TE-polarisation.
Figure~\ref{Fig:Fabry-Perot:L450:2d} shows the total transmittance  and the diffraction efficiency into the order $p = (-1,0)$,
for the high grazing incidence angle $\phi_0 = 89.999^{\circ}$.
The peak transmittance wavelengths for the TE and TM-polarisations are, respectively,
451.1~nm and 456.3~nm.

\begin{figure}[htbp]

\centerline{
\includegraphics[width=6.5cm]{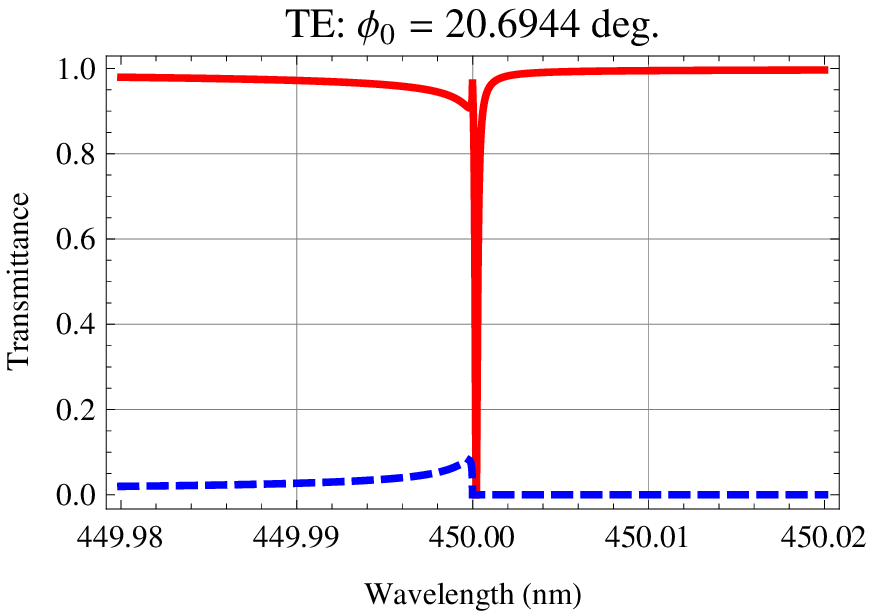}
}

\centerline{
\includegraphics[width=6.5cm]{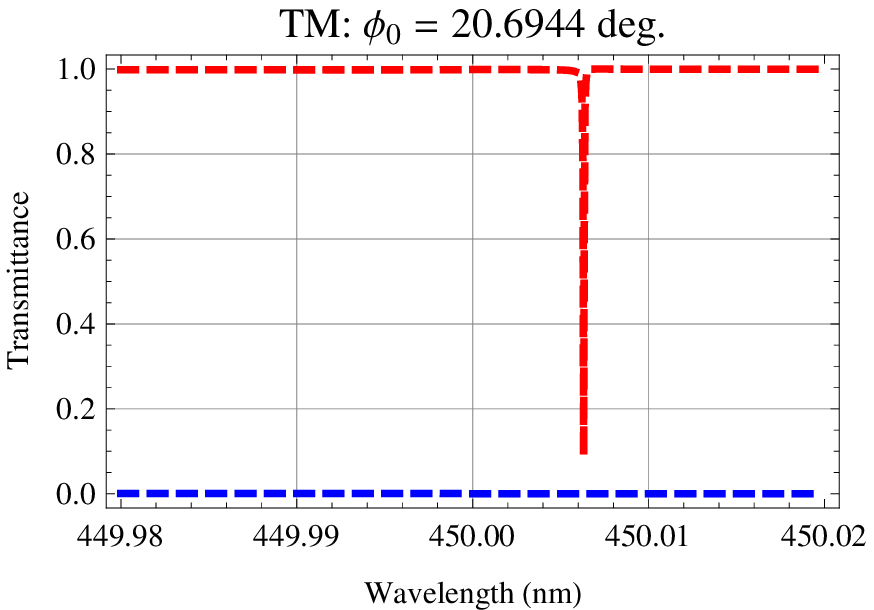}
}

\caption{Two-dimensional photonic crystal slab (Fabry-Perot resonance): 
The total transmittance (continuous red curves) and the diffraction efficiency into the order $p=(-1,0)$ 
(dashed blue curves). The wavelength for the Rayleigh anomaly at the order $p = (-1,0)$ is $\lambda = 450.0~{\rm nm}$.
}

\label{Fig2:Fabry-Perot:L450:2d}

\end{figure}

\begin{figure}[htbp]

\centerline{
\includegraphics[width=6.0cm]{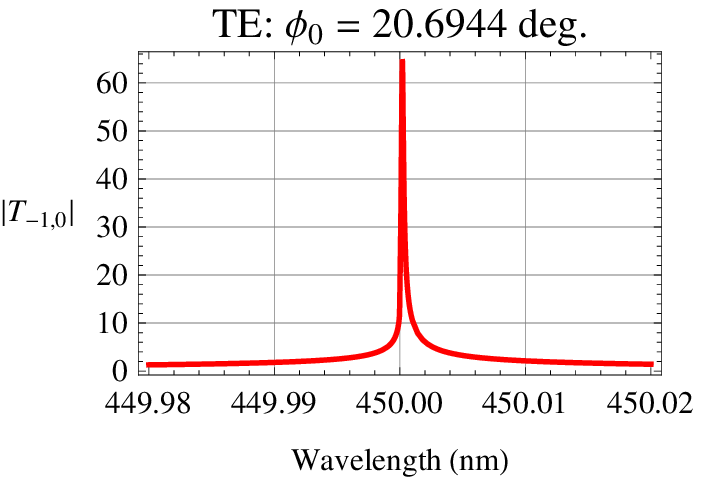}
}

\centerline{
\includegraphics[width=6.0cm]{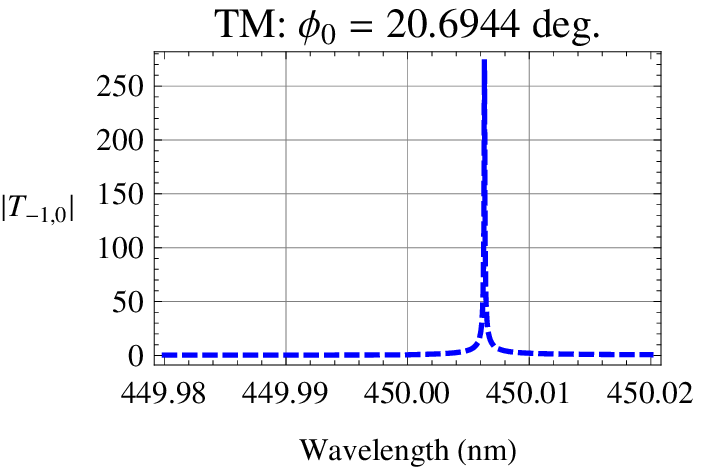}
}

\caption{Two-dimensional photonic crystal slab (field enhancement):
The transmission coefficient $|T_{-1,0}|$  corresponding to the first order diffraction efficiency 
shown in Fig.~\ref{Fig2:Fabry-Perot:L450:2d}.}

\label{Fig3:Fabry-Perot:L450:2d}

\end{figure}

\medskip
With the two-dimensional photonic crystal slab, we can also generate an enhanced diffracted field 
by combining the Fabry-Perot resonance with the Rayleigh anomaly.
A Rayleigh anomaly exists for the  order $p = (-1,0)$, near the wavelength $\lambda = 450~{\rm nm}$, 
when the angle of incidence is set to 
$\phi_0 = - \arcsin (\sin (89.999^{\circ}) - \lambda/(n_0 \, \Lambda)) = 20.6944^{\circ}$.
The slab thickness $h = 563~{\rm nm}$ is chosen such that the wavelength $\lambda = 450~{\rm nm}$ is close to a Fabry-Perot resonance
for the TE-polarisation.
We have plotted the total transmittance (continuous red curves) and the diffraction efficiency $e_{-1,0}$ into the order $p=(-1,0)$ 
(dashed blue curves) in Fig.~\ref{Fig2:Fabry-Perot:L450:2d} while the curves of the transmission coefficient $|T_{-1,0}|$
are shown in Fig.~\ref{Fig3:Fabry-Perot:L450:2d}.
We have a large transmission coefficient $|T_{-1,0}|$ for the TE-polarisation 
around the design wavelength $\lambda = 450~{\rm nm}$.
The peak transmission coefficient $|T_{-1,0}|$ for the TM-polarisation occurs in a wavelength band (centred around $\lambda = 450.006~{\rm nm}$)
where the order  $p=(-1,0)$ is evanescent.
This deviation from the operating wavelength $\lambda = 450~{\rm nm}$ 
can be explained by the fact the slab thickness $h = 563~{\rm nm}$ is not close enough to the calculated resonance thickness $h = 553.2~{\rm nm}$ for the TM-polarisation.
Actually, since the order $p=(-1,0)$ is evanescent at the wavelength $\lambda = 450.006~{\rm nm}$, the corresponding resonance thickness 
must be determined from the guided-mode resonance conditions Eqs.~(\ref{disersion:eq1:TE})-(\ref{disersion:eq1:TM}) for incidence by an evanescent plane wave.
By using  the average refractive index  $n_g = 1.3888$ and $n_g = 1.3908$, respectively, 
for the TE and TM-polarisations, and by applying Eqs.~(\ref{disersion:eq1:TE})-(\ref{disersion:eq1:TM}),
 we can obtain the following guided-mode  resonance thickness (at the wavelength $\lambda = 450.006~{\rm nm}$):
$h = 570.05~{\rm nm}$ for the TE-polarisation
and
$h = 561.1~{\rm nm}$ for the TM-polarisation.
Indeed, for the TM-polarisation the resonance thickness $h = 561.1~{\rm nm}$ is relatively close to the thickness $h = 563.0~{\rm nm}$
used for the calculations in Figs.~\ref{Fig2:Fabry-Perot:L450:2d} and \ref{Fig3:Fabry-Perot:L450:2d}
and this can explain why we have a guided-mode resonance for the TM-polarisation.

%
%%%%%%%%%%%%%%%%%%%%%%%
%
%

%

%
%%%%%%%%%%%%%%%%%%%%%%%
%

%
%%%%%%%%%%%%%%%%%%%%%%%
%

%
%%%%%%%%%%%%%%%%%%%%%%%
%

%
%%%%%%%%%%%%%%%%%%%%%%%%%%%%%%%%%%%%%%%%
%

\section{Potential applications}

\label{section:applications}

The results in Section~\ref{section:simulations} show that, 
when a non-specular order is resonantly coupled to a photonic crystal slab mode near a Rayleigh anomaly,
the corresponding diffraction coefficient can take very large values as the incident wavelength $\lambda$
approaches the Rayleigh anomaly wavelength.
This property can be applied to design photonic crystal slabs which can increase the intensity of an incident plane wave by a large factor.
The diffracted enhanced field is either propagating at a grazing angle or weakly evanescent.

\medskip
For most common gratings, the diffraction coefficient of a non-specular order $p$ typically takes very small values  
as its angle of diffraction approaches  $\pm\pi/2$.
When little energy is available, it is  not possible to fully take advantage of the fact that diffraction gratings can exhibit 
an arbitrarily large angular dispersion near a Rayleigh anomaly.
But, with the resonant cases illustrated in Figs.~\ref{Fig4:Fabry-Perot:L450} and \ref{Fig3:Fabry-Perot:L450:2d},
 it can make sense to consider the properties of a diffracted wave
in the limit where an incident wavelength is arbitrarily close to the Rayleigh anomaly wavelength.

\medskip
As an example, for a given incident ray and when a grating is rotating,
the angle of incidence $\phi_0$ of the ray at a time $t$
will change with an angular velocity denoted $\Omega_0$.
For a propagating diffraction order $p$, the diffracted wave will rotate
at an angular speed $\Omega_p = d \phi_p/ dt$.
By taking the time-derivative of the grating equation $\sin \phi_p = \sin \phi_0 + p \, \lambda / \Lambda$,
at a fixed wavelength,
the angular velocity $\Omega_p$ can be expressed in term of $\Omega_0$ 
as $\Omega_p = \Omega_0 \, \cos \phi_0 / \cos \phi_p$.
This shows that the angular velocity of a non-specular diffracted ray increases toward infinity as the angle of diffraction $\phi_p$ approaches  $\pm\pi/2$.
Such a fast moving ray can potentially shift the frequency of the incident light substantially.
The frequency shift can be analysed by using a Doppler shift formula 
since the diffracted light behaves like (or simulates) the light emitted from a surface which is rotating at the angular velocity $\Omega_p$~\cite{Dossou:AO:2016}.
Interestingly, diatoms can exhibit some oscillatory (or rotational) motion when subjected to water flows~\cite{Karp:LO:1998,Srajer:ABC:2009,Gutierrez:PB:2014},
which has some similarity with the tendency of plant leaves to oscillate under the wind.
Such an oscillation has the potential to strongly shift an incident frequency near a Rayleigh anomaly (when it is coupled with a Fabry-Perot resonance).
However, at this time, it is not possible to know if such a frequency shift is relevant to the photosynthetic processes in diatoms.
%

%
%%%%%%%%%%%%%%%%%%%%%%%%%%%%%%%%%%%%%%%%
%

\section{Conclusion}

Photonic crystal slabs can be designed to enhance the intensity of an incident plane wave by an extremely large factor.
Although we have only studied the case of a light transmission, the generalisation of the main results to a light reflection is straightforward.
We have given a clear physical explanation, based on the properties of the Fabry-Perot resonance and the Rayleigh anomaly, for the origin of this enhancement effect.
By applying the physical interpretations, we have developed an effective medium technique which can be used to efficiently design a photonic crystal slab,
with a resonance wavelength near a given value.
For the case of a scattering into a propagating diffracted order, the field magnification is due to a spatial compression of the incident wave 
and so the diffracted field propagates at a grazing angle of diffraction.
The numerical results also suggest that the photonic crystal slabs found in diatom frustules can produce a substantially enhanced field 
and it will be interesting to investigate the significance of this enhancement to the photosynthetic processes in diatoms.

\medskip
The combination of the Fabry-Perot resonance with the Rayleigh anomaly implies that a non-specular diffraction order can still transmit or reflect an intense light,
for incident wavelengths which are extremely close to the Rayleigh anomaly wavelength.
This opens up the possibility of an efficient operation of a diffraction grating 
in the limit where an incident wavelength is arbitrarily close to a Rayleigh anomaly wavelength,
a regime where diffraction gratings have an extremely large angular dispersion, e.g., with respect
to the incident wavelength or the incidence angle.
For an example, a rotation of the grating at a moderate angular speed can induce a diffracted field 
which rotates with an extremely large angular velocity,
and this in turn can produce a relatively large frequency shift.
%

%
%%%%%%%%%%%%%%%%%%%%%%%
%

\section*{Acknowledgments}
This research was supported by the Australian Research Council (ARC) 
(project number~CE110001018).
%

%
%%%%%%%%%%%%%%%%%%%%%%%%%%%%%%%%%%%%%%%%
%

%
%%%%%%%%%%%%%%%%%%%%%%%
%

% \section*{References}

%
%%%%%%%%%%%%%%%%%%%%%%%
%


%
%%%%%%%%%%%%%%%%%%%%%%%
%

%
\end{document}